\documentstyle[12pt,twoside,fleqn,espcrc1]{article}
\def\l{\left}
\def\r{\right}

\newcommand{\AmS}{{\protect\the\textfont2
  A\kern-.1667em\lower.5ex\hbox{M}\kern-.125emS}}

\title{ DEGENERACY SYMMETRY OF BARYON SPECTRA}

\author{ M. Kirchbach\address{Escuela de Fisica, 
        Universidad Autonoma de Zacatecas, \\ 
        A.\ P.\ C-600, ZAC-98062, Mexico}%
        }
       
\begin{document}

\maketitle

\begin{abstract}
We present a data-supported
new interpretation of the Rarita-Schwinger fields which reveals
a well-defined pattern in the masses, spins, and parities of the
N-, $\Delta$-, and $\Lambda$-hyperon excitations and which allows to 
probe the scale of the chiral phase transition for baryons. The 
degeneracy symmetry of the baryonic spectra emerges to be 
{\sc Isospin$\otimes$Space}-{\sc Time}.
\end{abstract}

\section{Introduction to the Relativistic Description of Higher-Spin 
States}
\subsection{Particle States and their Definition}
The definition of particle states is at the very heart 
of contemporary quantum theory of fields.
It takes its origin from the frame-independent 
Casimir invariants of the Poincar\'e group as
was first realized by Wigner in his work of late thirties \cite{Wi39}.
On that ground, quantum states of  free particles in a Poincar\'e
covariant framework are characterized by {\em mass} and {\em spin}, and 
transform for different inertial observers according to
\begin{equation}
|\psi'\, \rangle   = \exp \Big[ i 
( \epsilon^\mu P_\mu - \theta^{\mu \nu}I _{\mu \nu } )\,
\Big]\,   
|\psi\, \rangle \, .
\label{state}
\end{equation}
Here,  $P_\mu $ and the totally antisymmetric 
tensor $I_{\mu\nu}$ with $\mu (\nu)=0,1,2,3$ are
the generators of the Poincar\'e group which satisfy the commutation 
relations (Poincar\'e algebra):
\begin{eqnarray}
\lbrack I_{\mu\nu},I_{\rho \sigma }\rbrack &=&
i\left( g_{\nu\rho }I_{\mu\nu} - g_{\mu\nu }I_{\nu\sigma}
+g_{\mu\sigma}I_{\nu\rho} -g_{\nu\sigma}I_{\mu\rho}\right)\, ,\\
\lbrack P_\mu ,I_{\rho\sigma }\rbrack &=&i\left(
g_{\mu\nu} P_\sigma -g_{\mu\sigma}P_\rho\right)\, ,\quad 
\lbrack P_\mu,P_\nu\rbrack =0\, ,
\label{Po_alg}
\end{eqnarray}
where $g_{\mu\nu}$=diag$(1,-1,-1,-1)$ is the metric tensor, and
$\epsilon^\mu $ and $\theta^{\mu\nu}$ are continuous parameters.
{}Furthermore, $P_\mu$ generate translations,
$I_{kj}=\epsilon _{kjl}J_l$ with 
$k,(j,l)=1,2,3$ are the generators of rotations in the $kj$-plane, 
while $I_{0l}=K_l$  generate the boost along each of the $l$-axis.
Wigner postulated that particle states transform as 
{\it classical\/} unitary, and therefore infinite-dimensional, 
representations of the Poincar\'e group. Later on, Weinberg 
argued \cite{We64} that {\it quantized\/} non-unitary 
finite-dimensional representations can also be given particle 
interpretation, once unitarity of the corresponding
quantized field operators is ensured.

In the context of the existing baryonic spectra,
we shall establish the thesis that within the 
above-indicated framework one obtains not only unique spin
eigenstates, which can be identified with
particles, such as neutron, proton, etc., but also 
multi-spin mass-degenerate clusters. We will present a data-supported
new interpretation of the Rarita-Schwinger fields which reveals
a well-defined pattern in the masses, spins, and parities of the
nucleon-, $\Delta$-, and $\Lambda$-hyperon excitations 
and it allows to probe the scale of the chiral
phase transition for baryons.

\subsection{Description of Isolated High-Spin States }
It can be directly verified that the symmetric 
(right-handed) $\vec{R}=(\vec{J} +i\vec{K})/2$, and antisymmetric
(left-handed) $\vec {L}=(\vec{J} -i\vec{K})/2$ combinations of 
angular momentum, $\vec{J}$, 
and boost, $\vec{K}$, generate two independent left-handed (L) and
right-handed (R) su(2) algebras
\begin{equation}
\lbrack R_i,R_j\rbrack =i\epsilon_{ijk}R_k\, ,
\quad 
\lbrack L_i,L_j\rbrack =i\epsilon_{ijk}L_k\, ,\quad
\lbrack R_i,L_k\rbrack =0 \, .
\end{equation}
Obviously, it is possible to define the sets of right- and left-handed 
states, $|j_R,m_R\rangle$, and $|j_L,m_L\rangle $, respectively, such 
that
\begin{eqnarray}
R^2|j_R, m_R\rangle =j_R(j_R+1)|j_R, m_R\rangle\, ,
&\quad & R_z|j_Rm_R\rangle =m_R|j_R, m_R\rangle\, ,
\nonumber\\
L^2|j_L, m_L\rangle =j_L(j_L+1)|j_L, m_L\rangle\, ,
&\quad & L_z|j_L\, m_L \rangle =m_L|j_L, m_L\rangle\, .
\label{left_right}
\end{eqnarray}
Equations (\ref{left_right}) suggest to use $j_R$ and $j_L$ as labels 
for Lorentz multiplets according to $(j_R,j_L)$.
The dimensionality ($\mbox{dim}$) of $(j_R,j_L)$  
is now given by dim=$(2j_R+1)(2j_L+1)$. 
The most interesting $(j_R,j_L)$ states are those where either
$j_R$ or $j_L$ vanishes, i.e. the $(j_R,0)$ and 
$(0,j_L)$ states, and with $j_R=j_L:=j$. We denote those fields in turn by:
$
\Phi^R_{j}:=(j,0)$, and 
$\Phi^L_{j}:=(0,j)$.
As long as for such fields $R_i\Phi^L_{j}=L_i\Phi^R_{j}=0$ is true in
a spin-independent manner for all 
$\Phi^L_{j}$ and $\Phi^R_{j}$, the
boost generator turns out to
be particularly simple and related to $\vec{J}$ via 
\begin{equation}
{1\over 2}\left(\vec{J}+i\vec{K}\right)=0 
\,  \quad \mbox{for} \quad \Phi^L_{j}\, ,
\quad
{1\over 2}\left(\vec{J}-i\vec{K}\right)=0 \,  
\quad \mbox{for} \quad \Phi^R_{j}\, .
\label{L_r_boosts}
\end{equation} 
Consequently, $\vec{K}=i\vec{J}$ for $\Phi^L_{j}$, while
$\vec{K}=-i\vec{J}$ for $\Phi^R_{j}$. The momentum dependence of
the left-handed and right-handed fields is then determined through the
action of the boost onto the rest-frame fields as \cite{Ryder,DVA}
\begin{equation}
\Phi^R_{j} \left(\vec{p}\,\right) = \exp (
\vec{J}\cdot \hat{p}\, \varphi\,  )\,
\Phi^R_{j} \left(\vec{0}\, \right)\, , \quad 
\Phi^L_{j}\l(\vec{p}\, \r) = \exp \left(-\vec{J}\cdot \hat{p}\, \varphi\, 
\right)\,
\Phi^L_{j} \left(\vec{0}\, \right)\, .
\label{boosted_LR}
\end{equation}
Here, $\hat{p}$ stands for the boost direction, while $\varphi $ is the
boost parameter. 
As long as no skew sense can be defined at rest, 
$\Phi^L_{j}\l(\vec{0}\, \r)$ and $\Phi^R_{j}\l(\vec{0}\, \r)$
are indistinguishable (up to a phase factor that will be 
ignored for the time being \cite{DVA1}) and one can write
\begin{equation}
\Phi^L_{j}\l(\vec{0}\, \r)=\Phi^R_{j}\,\l(\vec{0}\, \r) \, .
\label{rest_fr}
\end{equation}
In inverting then, say, the first one of Eqs.~(\ref{boosted_LR}), 
one finds
\begin{eqnarray}
\Phi^R_{j}\,\l(\vec{0}\, \r) 
&=& \exp \l(-\vec{J}\cdot \hat{p}\, \varphi \r)\, 
\Phi^R_{j}\l(\vec{p}\, \r)
\label{R_L)ident}
\end{eqnarray}
which, after insertion into the second of the 
Eqs.~(\ref{boosted_LR}) yields the relation
\begin{equation}
\Phi^L_{j}\l(\vec{p}\, \r) =
\exp \l(-2\vec{J}\cdot \hat{p}\, \varphi\, \r)\, 
\Phi^R_{j}\l(\vec{p}\, \r)\, .
\label{L_R_coupl}
\end{equation}
Similarly, one finds
\begin{equation}
\Phi^{R}_j\l(\vec{p}\, \r) =
\exp \l( 2\vec{J}\cdot \hat{p}\, \varphi \, \r)\,
\Phi^L_{j}\l(\vec{p}\, \r)\, .
\label{R_L_coupl}
\end{equation}
The coupled equations (\ref{L_R_coupl}) and (\ref{R_L_coupl}) 
completely determine equation of motion and
propagation of the $(j,0)\oplus(0,j)$ fields. 
To be specific, we will 
illustrate this statement on the example of
the spin-1/2 field, where 
$\vec{J}={1\over 2}\, \vec{\sigma \, }$
with $\vec{\sigma\, }$ being the set of the well known three 
Pauli matrices. In this case, one has
\begin{eqnarray}
\Phi_{(1/2)}^R\l(\vec{p}\, \r)
&=& \exp \left( \vec{\sigma} \cdot \hat{p}\, \varphi \right) 
\, \Phi^L_{(1/2)}\l(\vec{p}\,\r)\,\, 
=\left(\cosh\left( \vec{\sigma}\cdot \hat{p}\, \varphi\right)  +
\sinh \left( \vec{\sigma}\cdot\hat{p}\, \varphi \right)
\right) \, \Phi^L_{\l(1/2\r)}\left( \vec{p}\, \right)\, ,
\nonumber\\
\Phi^L_{\l(1/2\r)}\left(\vec{p}\, \right)
&=& \exp \left( -\vec{\sigma}\cdot \hat{p}\varphi \right)\,
\Phi^R_{\l(1/2\r)}\left( \vec{p}\right)
=\left(\cosh
\left( \vec{\sigma}\cdot \hat{p}\, \varphi\right)  -
\sinh 
\left( \vec{\sigma}\cdot\hat{p}\, \varphi \right) \right)\,
\Phi^R_{(1/2)}\left( \vec{p}\, \right)\, .
\label{coupl_eqs}
\end{eqnarray} 
In using the Taylor-series expansion for the hyperbolic trigonometric 
functions and accounting for the circumstance that 
$\l(\vec{\sigma}\cdot \hat p\r)^2=1$, 
the equation set in (\ref{coupl_eqs}) takes the form
\begin{eqnarray}
\Phi^R_{\l(1/2\r)}\l(\vec{p}\, \r)&=&\l(\cosh\varphi +\vec{\sigma}
\cdot\hat{p}\,
\sinh\varphi\r)\,
\Phi^L_{\l(1/2\r)}\l(\vec{p}\, \r)\, ,\nonumber\\
\Phi^L_{(1/2)}(\vec{p}\, )&=&\l(\cosh\varphi -\vec{\sigma}
\cdot\hat{p}\, \sinh\varphi\r)\, 
\Phi^R_{\l(1/2\r)}\l(\vec{p}\, \r)\, .
\label{pre_Dirac}
\end{eqnarray}
{}Finally, insertion of the relations,
$
\cosh \varphi ={p_0/ m}$,
$
\sinh\varphi ={|\vec{p}\,|/ m}
$,
into Eqs.~(\ref{pre_Dirac}) leads to
\begin{eqnarray}
m\, \Phi^R_{\l(1/2\r)}\l(\vec{p\, }\r)&=&\l(p_0+\vec{\sigma}\cdot
\hat{p\, }\, |\vec{p\, }\, |\, \r)\, 
\Phi^L_{(1/2)}(\vec{p\, })\, ,\nonumber\\
m\, \Phi^L_{(1/2)}\l(\vec{p\, }\r)&=&\l(p_0-\vec{\sigma}\cdot\hat{p\, }\, 
|\vec{p\, }\, |\, \r)\, 
\Phi^R_{\l(1/2\r)}(\vec{p\, })\, .
\label{Weyl_eqs}
\end{eqnarray}
This set of equations can be cast into the matrix form
\begin{equation}
\left(\begin{array}{cc}
-m&p_0+\vec{\sigma\, }\cdot\vec{p\, }\\
p_0-\vec{\sigma\, }\cdot\vec{p\, }&-m
\end{array}\right)
\left(\begin{array}{c}
\Phi^R_{\l(1/2\r)}(\vec{p\, })\\
\Phi^L_{\l(1/2\r)}\l(\vec{p\, }\r)
\end{array}\right)=0\, ,
\label{Matrix_form}
\end{equation} 
which is nothing but the Dirac equation, $\l(\gamma^\mu p_\mu-m\r)
\psi\l(p\r)=0$, in the Weyl representation.
Equations ~(\ref{pre_Dirac}) and (\ref{Weyl_eqs}) show
that  the {\em linearity} of the Dirac equation is due to the 
fact that $\vec{\sigma\, }^2=1$.
In general \cite{DVA,DVA1}, the order of the differential equations
for spin-$j$ particles is $\partial_\mu ^{2j}$ 
and is related to the fact that $p^{2j}_\mu $ is the highest power of
the momentum that appears in the expansion of 
$\exp ( 2\vec{J}\, \cdot\vec{\phi }\, )$. 
The general consequence is that the equation of motion for 
single higher-spin states are inevitably higher-order differential 
equations. Now, in what follows, we shall focus onto
the properties of the boost generator and reveal need for an
inherent handedness of the Lorentz multiplets. 

Due to the non-vanishing commutator $\lbrack J_i, K_j\rbrack \not= 0$,
the boost is in general non-diagonal in the quantum number of spin. 
In addition, in transforming as an ordinary polar vector, the boost
behaves \cite{KimNoz} as a 1st-rank spherical tensor, 
$Y_{1m}\l(\hat{p\, }\r)$, and connects opposite parity states with spins 
differing by either zero, or, by one unit. The non--vanishing matrix 
elements of $K_z$ are determined as 
\begin{eqnarray}
\Big
\langle j^{-\pi },m\Big\vert K_z\Big\vert j^\pi, m \Big\rangle 
=\lambda_j m\, ,
&\quad & \Big\langle \l(j^{-\pi} -1\r), m\Big\vert K_z
\Big\vert j^\pi, m\Big\rangle =
\zeta_j \sqrt{j^2 -m^2}\, ,
\label{bst_mael}
\end{eqnarray}
while those of the ladder boost-generators, 
$K_\pm =K_x\pm iK_y$, are given by
\begin{eqnarray}
\langle j^{-\pi },m|K_\pm |j^\pi , m\mp 1\rangle &=&\lambda_j
\sqrt{\l(j\pm m\r)\l(j \mp m+1\r)}\, , \nonumber\\
\langle j^{-\pi } -1,m|K_\pm |j^\pi , m\mp 1\rangle &= &\pm \zeta_j
\sqrt{\l(j \mp m\r)\l(j \mp m+1\r)}\, .
\label{ladder-boost}
\end{eqnarray} 
Here $\pi $ stands for the parity,
and $\lambda_j$ and $\zeta_j$ are parameters defined as
\begin{equation}
\lambda_j=i{
{ j_{min} \lambda } \over { j\l(j+1\r)}
}\, ,
\quad \zeta_j^2= {
{ \l(j^2-j_{min}^2\r)\l(j^2-\lambda^2\r) }\over
{j^2\l(4j^2-1\r)}
}\, ,\quad \lambda =\pm \l(j_{max} +1\r)\, ,
\label{laj_dzj}
\end{equation}
where $j_{min}$ and $j_{max}$ are in turn the
minimal and maximal spin within the representation considered.
To be specific, for the right-handed $\l({1\over 2},0\r)$ 
state one has $\lambda ={3\over 2}$ while for the left-handed
$\l(0,{1\over 2}\r)$ state $\lambda =-{3\over 2}$. 

Equation~(\ref{bst_mael}) shows that if the 
boost is to be diagonal in spinor space and have non-vanishing
matrix elements of the type 
$\langle \Phi^{R\l(L\r)}_j\l(\vec{p\, }\r)\, |K_z
|\Phi^{R \l(L\r)}_j\l(\vec{p\, }\r) \rangle $, 
the spinors $\Phi ^{R\l(L\r)}_j\l(\vec{p\, }\, \r)$ have to be 
parity mixed states. This is the meaning of ``handedness''.
As an example one may think of the 
$\l(1,0\r)\oplus\l(0,1\r)$ multiplet that can be associated 
with  the electromagnetic field strength tensor in the massless limit. 
This multiplet has the symmetric $\l(\vec{B}+i\vec{E}\r)$ and antisymmetric
$\l(\vec{B}-i\vec{E}\r)$ combinations of the vector electric ($\vec{E}\, $) 
and pseudovector magnetic ($\vec{B}\, )$ fields as underlying right-handed 
$\l(1,0\r)$- and left-handed $\l(0,1\r)$ spin-1 states, respectively. 
The lessons to be learned from the above considerations are two: 
(a) Classical single-spin-j Lorentz multiplets carry an inherent 
handedness that requires two states of opposite parities as 
ingredients, and (b) 
Such states are described by partial differential equations 
of the order $\partial_\mu ^{2j}$.

\subsection{Rarita-Schwinger Fields as Multi-Fermion Clusters}
A method for treating higher-spin states
that is different from the one presented in the previous section
was initiated by Weinberg around mid sixties \cite{We64}. 
Weinberg suggested to consider linear differential equations
for finite-dimensional non-unitary representations of the Lorentz group 
that are described in terms of totally symmetric traceless
rank-$k$ Lorentz tensors with Dirac spinor components, i.e. fields of 
the type
\begin{equation}
\Psi_{\mu_1\mu_2...\mu_k}:=\left({k\over 2},{k\over 2}\right)\otimes
\left[\left({1\over 2},0\right)\oplus \left(0,{1\over 2}\right)\right]\, .
\label{RS_fields}
\end{equation}
These fields satisfy the Dirac equation for each Lorentz index
\begin{eqnarray}	
\l(i\partial_\lambda \gamma^\lambda - M\r)\Psi_{\mu_1 \mu_2
\cdots \mu_k} = 0\, .
\label{Dirac_Proca}
\end{eqnarray}
The simplest field of the type in Eq.~(\ref{RS_fields}), 
with $k=1$, was considered more than two decades 
earlier by Rarita and Schwinger \cite{RS} 
and applied to the description of a spin-3/2 
particle. In general, all fields of the type in (\ref{RS_fields}) 
will be referred to as Rarita-Schwinger (RS) fields.
The RS fields are conglomerations of fermions of different spins and 
parities. This statement is best illustrated at the level of the 
group O(4), the Wick rotated compact form of the Lorentz group.
In noticing that the bosonic as well as the fermionic part of
the RS field are constructed out of parity-mixed fields, one can
pick up and consider for the sake of concreteness, the coupling of,
say, a positive parity Dirac fermion to an O(4) boson composed of 
O(3) states of either natural $(\eta =+)$, or, 
unnatural $(\eta =- $) parities. 
These O(3) states, which can be viewed as mass degenerate,
carry all integer internal angular momenta, $l$, with 
$l=0,\dots,\sigma-1$ and transform with 
respect to the space inversion operation ${\cal P}$ as
\begin{eqnarray}	
{\cal P} \sigma _{\eta ;lm} = \eta e^{i\pi l}\sigma_{\eta;l-m}\,, 
\quad
l^P=0^\eta,1^{-\eta},\dots,\l(\sigma -1\r)^{-\eta}\, , \quad 
m=-l,\dots,l\, .
\label{parity}
\end{eqnarray}
In coupling now the positive-parity part of the Dirac spinor 
to $\l({k\over 2},{k\over 2}\r)$ from above, the spin ($J$) and
parity $(P$) quantum numbers of the baryon resonances are created as
\begin{eqnarray}	
J^P = {1\over 2}^\eta,{1\over 2}^{-\eta}, {3\over 2}^{-\eta}, \dots,
\bigg(k+{1\over 2}\bigg)^{-\eta} \, , \quad k=\sigma -1\, .
\label{coupl_scheme}
\end{eqnarray}
In the following, we will use for  
the spin-sequence in Eq.~(\ref{coupl_scheme})
the short-hand notation $\sigma_ {2I, \eta}$, with 
$I$ standing for the isospin of the states considered,
and $\sigma =k+1$.
Equation (\ref{parity}) shows that $\sigma $ directly
relates to the principal quantum number of the Coulomb problem.
The states of different 
spins and opposite parities that accompany the highest-spin state 
of any $\l({k\over 2}, {k\over 2}\r)$-, and, consequently, of any  
RS field, serve an important purpose. They prevent that the boost 
generator vanishes within the multiplet under consideration.
To illustrate this statement, let us consider as a simple 
example the bosonic part $\l({1\over 2}, {1\over 2}\r)$ of the  
RS field with $k=1$ which is the source of the spin-$1/2^\pm $ 
companions of the spin-$3/2^-$ field.
{}From Eq.~(\ref{bst_mael}) follows that 
the boost generator $K_z$ mixes the O(3) scalar and vector 
subspaces of $\l({1\over 2},{1\over 2}\r)$ as 
the only non-vanishing matrix elements of $K_z$ in this multiplet
are   
\begin{eqnarray}
\langle j^\prime =0^+, m=0| K_z|
j=1^-, m= 0\rangle &=& \zeta_1\sqrt{1^2-0^2}\, ,\nonumber\\
\langle j^\prime = 1^-,m=0|K_z|j=0^+,m= 0\rangle &=&
\zeta_1\sqrt{1^2 -0^2}\, .
\label{bz_mae}
\end{eqnarray}
{}From this, and $\lambda =2$, the parameter $\zeta_1$
is obtained from Eq.~(\ref{laj_dzj}) as $\zeta_1=i$. 
In this way the standard representation of the  $K_z$ matrix in the 
$\l(x_0, \vec{x}\r)$ space is obtained as
\begin{eqnarray}
K_z&=&\left(\begin{array}{cccc}
0&0&0&i\\
0&0&0&0\\
0&0&0&0\\
i&0&0&0
\end{array}\right)\, .
\label{b_z}
\end{eqnarray}
Since Lorentz and spinor indices of the RS fields
always factorize, the boost of this 
representation space is most naturally obtained in boosting 
its bosonic part independently from the fermionic one.
Equations (\ref{bst_mael}) and (\ref{ladder-boost}), 
in combination with Eqs.~(\ref{parity}) and (\ref{coupl_scheme}),
clearly illustrate that the presence of the lower-spin components of a 
RS cluster guarantee in a simple manner its covariance with respect to 
Lorentz transformations. Note that for RS-, as well as 
$\l(j,0\r)\oplus\l(0,j\r)$ fields, 
both $K_z$ and $K_\pm $ are non-vanishing.

Now, Weinberg's suggestion was to consider the lower-spin components 
of the RS fields as redundant, unphysical states that have to be 
projected out by a set of suitably chosen Lorentz
covariant auxiliary conditions.
In doing so, one eventually would describe a baryon of 
single spin-$J=k+{1\over 2}$ and  {\it fixed parity\/}
as the highest-spin state of the $\Psi_{\mu_1\mu_2...\mu_k}$ field
and use the linear differential equation (\ref{RS_fields}). 

When applied to the  $({1\over 2},{1\over 2})\otimes
\lbrack ({1\over 2},0)\oplus (0,{1\over 2})\rbrack $ field, 
the auxiliary conditions eliminate eight degrees of freedom, 
usually associated with the lower spin-$1/2^+ $ and spin-$1/2^-$ 
components from Eq.~(\ref{coupl_scheme}).
Immediately, one notes that such a description is not
necessarily compatible with
the general nature of the boost generator in Eqs.~(\ref{bst_mael})
and (\ref{ladder-boost}).  
It is little wonder, that truncated RS fields suffer various pathologies 
(see \cite{VeZw} for details).
So one has to face the unavoidable conclusion of the principal 
difficulty to describe isolated higher-spin states of
fixed parity in terms of linear differential equations.
Note, that the problem arose out of the tacit assumption that
Nature produces baryons as isolated higher-spin states. 
A glance at the baryon spectra teaches us that actually nature strongly
favors the excitations of complete RS clusters over that of 
isolated higher-spin states.

We now show that the new interpretation of the RS fields is supported
by the data on baryonic spectra.

\section{Light-Quark Baryons as Rarita-Schwinger Clusters} 

One of the basic quality tests for any model of composite baryons is
the level of \hbox{accuracy} reached in describing the 
nucleon and $\Delta $ excitation spectra. 
In that respect, the knowledge on the degeneracy group of 
baryon spectra appears as a key tool in constructing the underlying 
strong-interaction dynamics.
To uncover it, one has first to analyze isospin by isospin how the 
masses of the resonances from the full listing in Ref.~\cite{Part} 
spread with spin and parity. Such an analysis has been performed in 
our previous work \cite{Ki97-98a} where it was found that 
Breit-Wigner masses reveal on the $M/J$ plane a 
well pronounced spin- and parity clustering. There, it was further shown 
that the quantum numbers of the resonances belonging to a particular 
cluster fit into Lorentz group representations of the RS  type. To be
specific, one finds the three RS clusters with $k=1,3$, and $5$ in
both the nucleon $(N)$ and $\Delta$ spectra. 
In terms of the notations introduced above, all reported
light-quark baryons with masses \hbox{below} 2500~MeV (up to
the $\Delta \l(1600\r)$ resonance that is most probably an
independent hybrid state), 
have been shown in Ref.~\cite{Ki97-98a}  
to be completely accommodated by the RS clusters $2_{2I,+} $, 
$4_{2I,-}$, and $6_{2I,-}$, having states of highest 
spin-$3/2^-$, $7/2^+$, and $11/2^+$, respectively (see Fig.~1). 
In each one of the nucleon, $\Delta$, and $\Lambda$ hyperon spectra, 
the natural parity cluster $2_{2I,+} $ is always of lowest mass. 
It is considered to reside in a Fock space, ${\cal F}_+$, 
built on top of a scalar vacuum. From Eqs.~(\ref{parity}) and 
(\ref{coupl_scheme}) follows that
the $2_{2I,+}$ clusters, where $I=1/2, 3/2$, and $0$, always unite the
first spin-${1\over 2}^+$, ${1\over 2}^-$, and ${3\over 2}^-$
resonances. {}For the non-strange baryons, $2_{2I,+}$
is followed by the unnatural parity clusters
$4_{2I,-}$, and $6_{2I,-}$, which we view to reside in a 
different Fock space, ${\cal F}_-$, that is built on top of a 
pseudoscalar vacuum. 
To be specific, one finds all the seven 
$\Delta$-baryon resonances $S_{31}$, $P_{31}$, $P_{33}$, $D_{33}$, 
$D_{35}$, $F_{35}$ and $F_{37}$ from $4_{3,-}$ to be squeezed within 
the narrow mass region from 1900~MeV to 1950~MeV, while the $I=1/2$ 
resonances paralleling them, of which only the $F_{17}$ state is still
``missing'' from the data, are located around
1700$^{+20}_{-50}$~MeV (see Fig.~1). Therefore, the F$_{17}$ resonance 
is the only non-strange state with a mass below 2000 MeV which 
is ``missing'' in the present RS classification scheme.
Further paralleling baryons from the third nucleon and
$\Delta$ clusters with $\sigma=6$, one finds in addition the four
states $H_{1, 11}$, $P_{31}$, $P_{33}$, and $D_{33}$ with masses above
2000~MeV to be ``missing'' for the completeness of the new
classification scheme. The $H_{1, 11}$ state is needed to parallel the
well established $H_{3, 11}$ baryon, while the $\Delta$-states
$P_{31}$, $P_{33}$, and $D_{33}$ are required as partners to the (less
established) $P_{11}$(2100), $P_{13}$(1900), and $D_{13}$(2080)
nucleon resonances. 
{}For $\Lambda $ hyperons, sparse data prevent a conclusive analyses.
Even so, see Fig.~2, 
the RS pattern is already apparent in the reported spectrum.  
\begin{figure}[htb]
\vskip 5.0cm
\includegraphics{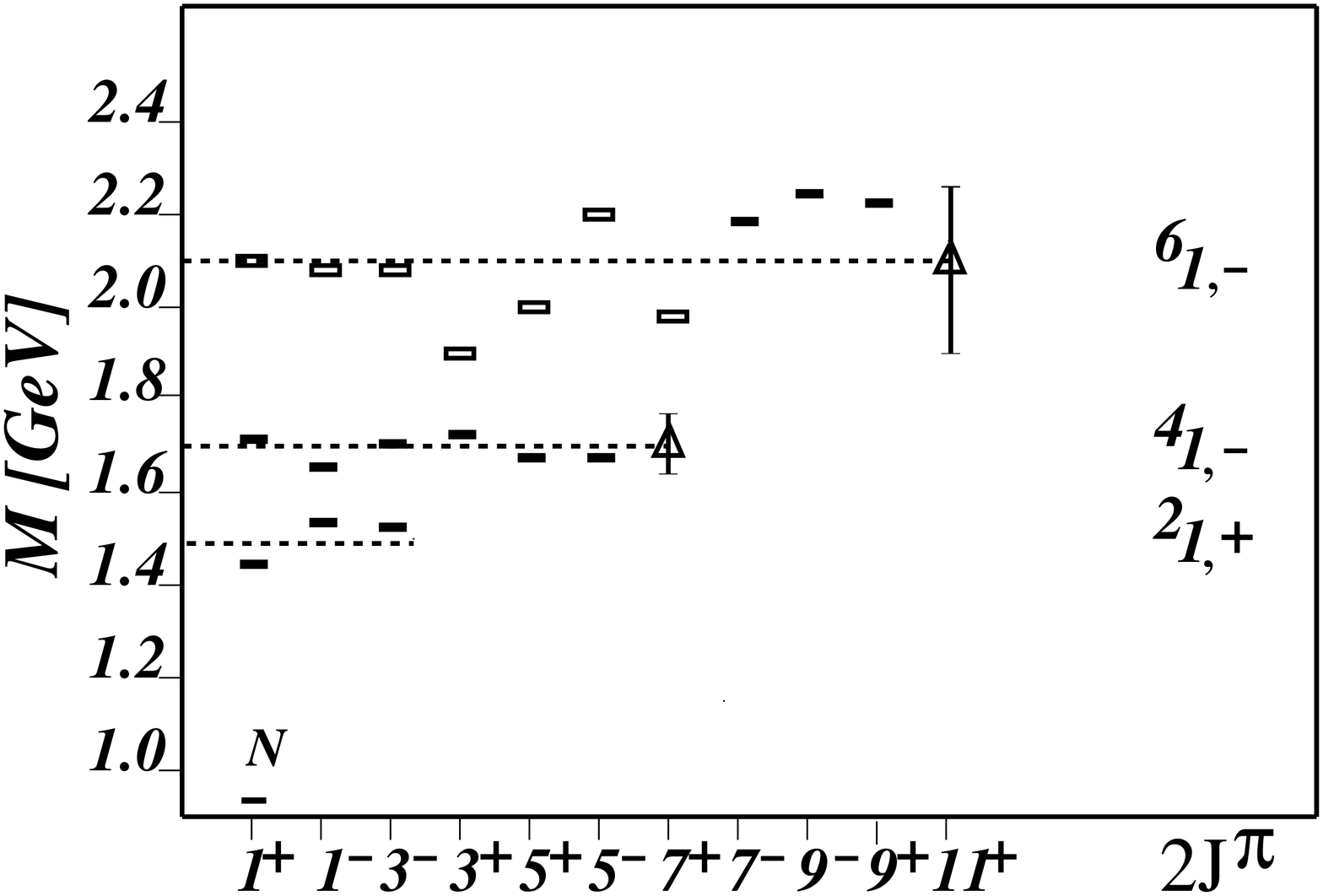}
\includegraphics{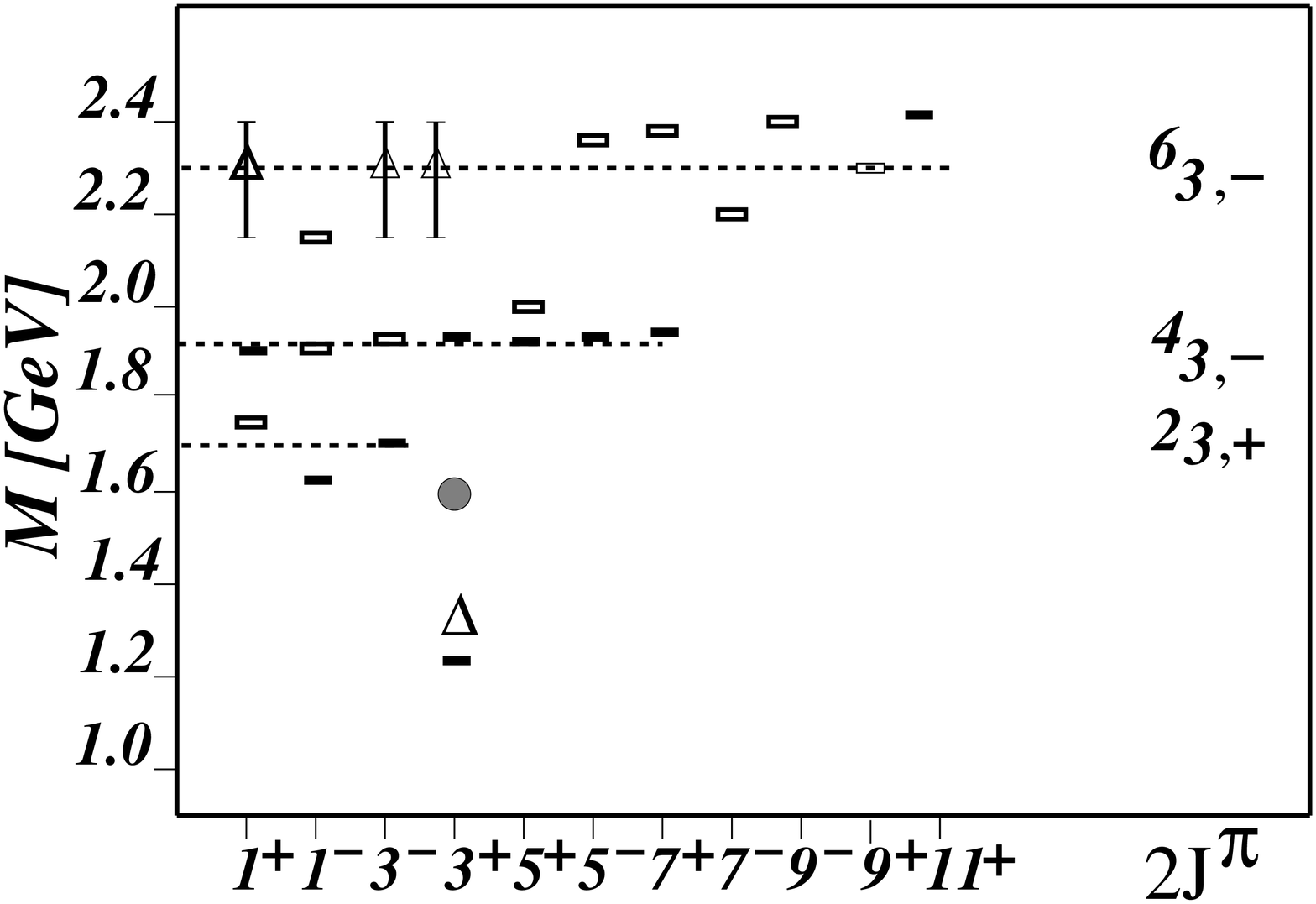}
\caption{Rarita-Schwinger clustering of light-quark resonances.
The full bricks stand for three-to five-star resonances, the empty bricks
are one- to two-star states, while the triangles represent states that are
``missing'' for the completeness of the three RS clusters.
Note that ``missing''  $F_{17}$ and $H_{1,11}$ nucleon excitations 
(left figure) appear as  four-star resonances in the 
$\Delta $ spectrum (right figure). 
The ``missing'' $\Delta $ excitations $P_{31}$, $P_{33}$, and $D_{33}$
from $6_{3,-}$ are one-to two star resonances in the nucleon counterpart
$6_{1,-}$. The $\Delta (1600)$ resonance (shadowed oval) drops 
out of our RS cluster systematics and we view it
as an independent hybrid state.}
\label{fig:RS_degeneracy}
\end{figure}
The (approximate) degeneracy group of baryon spectra 
as already suggested in Refs.~\cite{Ki97-98a} 
and further justified here, is, therefore, found to be:
$SU(2)_I\otimes O(1,3)_{ls}$,
i.e., Isospin$\otimes$Space-Time symmetry.

Within our scheme, the inter-cluster spacing of $200$ to $300$~MeV 
is larger by a factor of $3$ to $6$ as compared
to the mass spread within the clusters. For example, the $2_{1,+}$,
$2_{3,+}$, $4_{1,-}$, and $4_{3,-}$ clusters carry
the maximal mass splitting of $50$ to $70$~MeV.

\noindent
The above considerations establish that:
\begin{enumerate}
\item  Observed excited light-flavor baryons
preferably exist as multi-fermion clusters
that are described 
in terms of the three RS multiplets $2_{2I,+}$, $4_{2I,-}$ and 
$6_{2I,-}$.
\item The above RS clusters  exhaust all the resonances
observed in the $\pi N$ scattering channel 
(up to the $\Delta \l(1600\r)$ state). They 
 constitute an almost accomplished excitation mode in its own rights,
as only 5 resonances are ``missing'' for the completeness of
this structure.
\item As long as the internal parity of the clusters changes
from natural for the first one, to unnatural, for the subsequent two,
the question arises whether this change signals a phase transition for
baryons. Next section is devoted to this problem.
\end{enumerate}

\section{Probing
the Scale of the Chiral Phase Transition by ``Missing'' Resonances}
A conspicuous feature of the RS cluster excitation mode
are the even $\sigma $-values and 
the parity change when going from the first, $2_{2I,+}$, 
to the second and third RS clusters
$4_{2I,-}$, and $6_{2I,-}$, respectively. In that regard, the
absence of odd-$\sigma $ clusters with parity that is
opposite to the one of the observed even-$\sigma $ clusters,
needs explanation. A decomposition into RS clusters of the 
three-quark Hilbert space, ${\cal H}^{3q}$, with one excited 
quark at the $1s-1p-...-1f-1g$ orbits, revealed the principal
possibility \cite{MK2000} of having as an additional baryon 
excitation mode the  $3_{2I,-}$, $3_{2I,+}$ and $5_{2I,+}$ RS 
cluster sequence with highest spins $5/2^+$, $5/2^-$, and $9/2^+$, 
respectively. These resonance groups are characterized by
parities that are opposite to the observed ones. Their
presence or absence probes the scale of
the chiral symmetry realization in the light-quark baryon 
spectra. Indeed, in case the Nambu-Goldstone mode of chiral
symmetry that selects ${\cal F}_+$ and natural parities, 
would extend for, say, the nucleon, to $\sim $1500 MeV, one can not
expect to find the unnatural parity $3_{1,-}$ cluster in that
excitation domain. Above 1600 MeV, ${\cal F}_+$ changes to 
${\cal F}_ -$. In case ${\cal F}_-$ has been selected by
a hidden mode of chiral symmetry, no natural parity clusters can occur 
above 1600 MeV and $3_{2I,+}$ and $5_{2I,+}$ will be absent there.
If, however, chiral symmetry above 1600 MeV is realized  in the manifest 
Wigner-Weyl mode, then  the natural parity clusters $3_{1,+}$ and 
$5_{1,+}$ will show up. Whatever the chiral symmetry mode above
1600 MeV may be, the data strongly hint onto a parity change of the 
hadronic vacuum between 1550 MeV and 1600 MeV for nucleons.
Therefore, in that mass region one may expect
at least a meta-stable co-existence of two parity degenerate vacua
and there the occurrence of parity doublets can be expected.
On the grounds of the analyses performed in Ref.~\cite{MK2000}
one expects a cluster that is described in terms of
a totally antisymmetric Lorentz tensor with Dirac spinor
components. Such a cluster gives rise to spin-$1/2^\pm$ and 
spin-$3/2^\pm $ parity doublets. 
The scale of the chiral phase transition for nucleons is therefore
predicted to be around 1600 MeV.
This means that opposite to the considerations of Ref.~\cite{Oka},
it is the $S_{11}(1650)$ resonance that, in being
built on top of a $0^-$ internal boson degree of freedom,
has to be considered as the ``would be'' parity partner of the 
nucleon and has to
enter the linear sigma model with parity doublets.
The S$_{11}(1535)$ resonance used instead in \cite{Oka}
has $1^-$ as internal orbital angular momentum and does not 
match with the scalar internal mode of the nucleon. 
In the context of the classification of baryons as RS clusters,
states like the  D$_{13}$(1700) and P$_{13}$(1720),
or, the $D_{15}(1675)$ and $F_{15}(1680)$ resonances, 
do not pair, because, in belonging to the same RS cluster,
their internal orbital angular momenta
differ by one unit, instead of being equal but of opposite parities.

\section{Empirical Mass Formula for the Rarita-Schwinger Clusters}
The reported mass averages of the resonances from the 
RS multiplets with $k=1,3$, and $5$ are well described by means of the
following simple {\it empirical\/} recursive relation:
\begin{eqnarray}	
M_{\sigma'} - M_{\sigma} = m_1\bigg({1\over {\sigma^2}} - 
{1\over {(\sigma')^2}}\bigg) + {1\over 2} m_2
\bigg({{\sigma '^2-1}\over 2}- {{\sigma ^2 -1}\over 2}\bigg)\, ,
\label{Balmer_ser}
\end{eqnarray}
where, again, $\sigma=k+1$.
The two mass parameters take the values $m_1=600$~MeV, and
$m_2=70$~MeV, respectively. The first term on the r.h.s. in
Eq.~(\ref{Balmer_ser}) is the typical difference between the energies
of two single particle states of principal quantum
numbers $\sigma $, and $\sigma '$, respectively, occupied by a particle 
with mass $m$ moving in a Coulomb-like potential of strength $\alpha_C$ 
with $m_1=\alpha_C^2m/2 $. The term
\begin{equation}	
{{\sigma ^2-1}\over 2}=2 {k\over 2}\bigg({k\over 2}+1\bigg)\, ,
\quad \mbox{with} \quad k=\sigma -1 \, ,
\label{I(I+1)_O(4)}
\end{equation}
in Eq.~(\ref{Balmer_ser}) is the generalization of the
three-dimensional $j(j+1)$ rule (with $j=k/2$) to four Euclidean
dimensions \cite{KimNoz} and describes a generalized O(4) {\it
rotational band\/} (see Fig.~2).
\begin{figure}[htb]
\vskip 5.0cm
\includegraphics{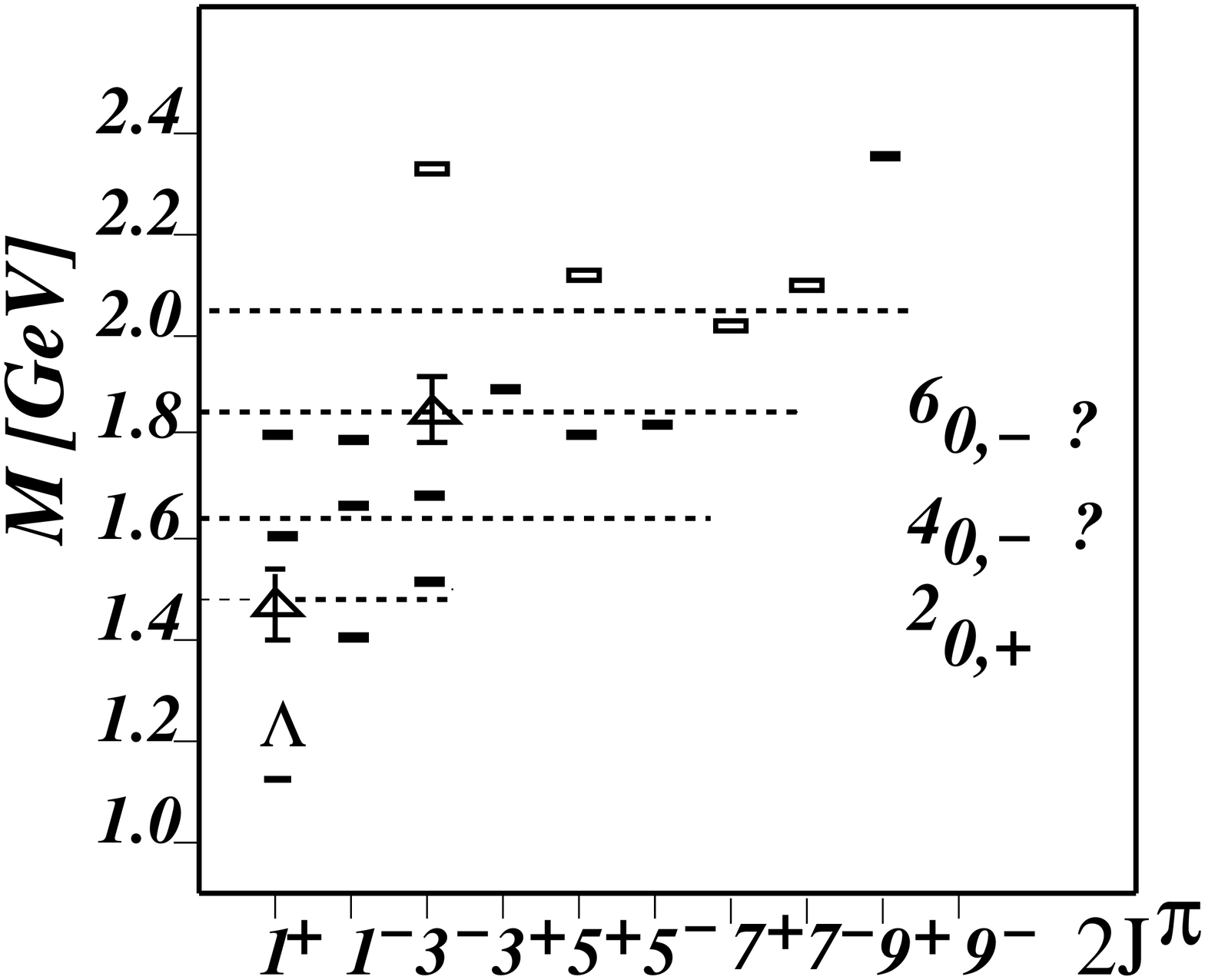}
\includegraphics{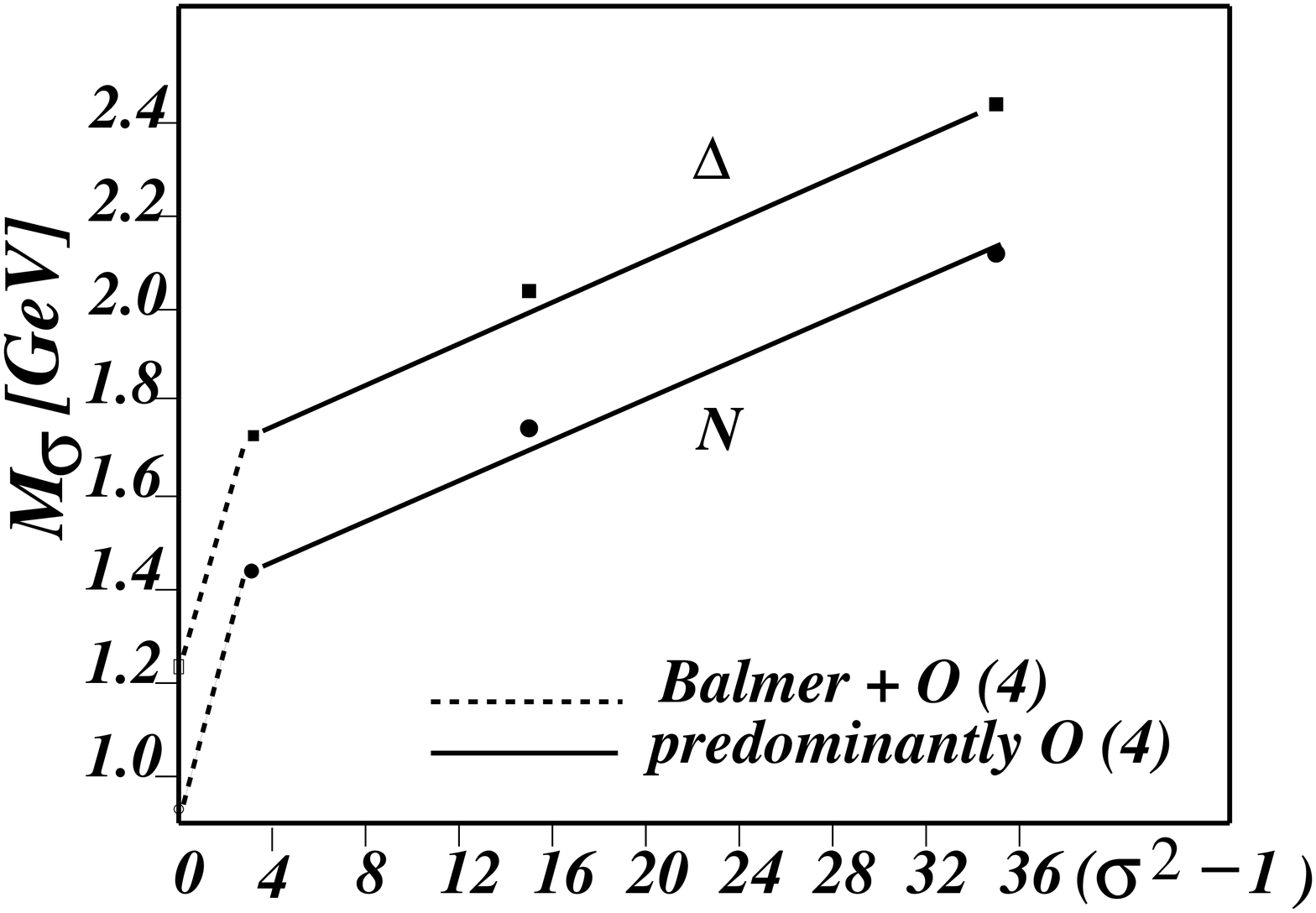}
\caption{Clustering in the $\Lambda $ hyperon spectrum (left). 
O(4) rotational bands of nucleon (N) and $(\Delta )$ excitations
(right). Notations as in Fig.~1. }
\label{fig:O4_band}
\end{figure}
The parameter $1/m_2=2. 82$~fm corresponds to the
moment of inertia ${\cal J}=2/5 MR^2$ of some ``effective'' rigid-body
resonance with mass $M =1085$~MeV and a radius $R=1.13$~fm.
Note that while the splitting between the Coulomb-like states
decreases with increasing  $\sigma $, the
difference between the energies of the rotational states increases
linearly with $\sigma $ so that the net effect is a slightly increasing 
spacing between the baryon cluster levels (see Fig.~2).

\section{Summary and Outlook}
We conclude that the lower-spin components of the Rarita-Schwinger
fields are forced upon by the properties of the Lorentz boost
and should not be projected out.  
These fields  are realized in the spectra of the light-flavor
baryons, where the landscape of the excitations
is well structured along the RS classification scheme,
rather than populated  by randomly distributed resonances.
The (approximate) degeneracy symmetry turned
out to be SU(2)$_I\otimes $O(1,3)$_{ls}$.
Within this context, the $2_{2I,+}$, $4_{2I,-}$, and
$6_{2I,-}$ RS clusters observed so far in the $\pi N$ scattering
channel by the LAMPF at LANL, constitute an almost accomplished 
excitation mode with only five states ``missing''. We further showed 
how presence or absence of the independent cluster
excitation sequence $3_{2I,-}$, $3_{2I,+}$, and $5_{2I,+}$, probes
the scale of the chiral phase transition for baryons.  
In this way, the ``missing'' state search
program through the CLAS collaboration at JLAB \cite{Burkert} obtains
an additional motivation, that is conceptually different from the 
SU(6)$_{SF}\otimes$ O(3)$_L$ classification scheme.

The dynamical origin for the RS clustering is still lacking 
a unique explanation and it remains a challenge for future research.
The factorization of isospin from the space-time symmetry
in SU(2)$_I\otimes $O(1,3)$_{ls}$ is strongly supported by 
QCD, where the isodoublet light quarks and the isosinglet
heavy-flavor quarks are the established isospin degrees of freedom.
On the other hand, any QCD solution has necessarily to be
a Lorentz covariant object.
In Ref.~ \cite{MK2000} the bosonic parts of the Lorentz clusters, 
such as  $({k\over 2}, {k\over 2})$ from the RS field in 
Eq.~(\ref{RS_fields}), were considered as independent fundamental
bosonic degree of freedom of baryon structure, and 
the term ``hyperquark'' was coined for them. There, the clustering was
modeled after a O(4) invariant quark-hyperquark correlation.  
{}From a slightly different QCD perspective, the clusters can also 
be viewed as string solutions associated with a linear action. 
Such strings are also complex
multi-fermion systems having the Dirac state as a limit \cite{Savvidy}. 
They can be described in terms of a multi-dimensional  
Lorentz invariant field $\Psi $ that satisfies the
Dirac-like equation $ (\Gamma^\mu p_\mu -M)\Psi =0\,$, where
Lorentz invariance imposes certain condition onto the 
$\Gamma$ matrices. The consideration presented above show 
that the RS fields naturally fit into this QCD string scheme.
In Ref.~\cite{Savvidy}, a solution with a RS-like degeneracy was 
reported. 
On the whole, our view is that a structured baryon spectrum that 
shares common flavor and relativistic symmetries with QCD is more 
likely to be linked via an appropriate effective theory
to first principles of strong-interaction dynamics than 
a spectrum with states distributed at random.

\section*{Acknowledgment}
Extensive and supportive discussions with D. V. Ahluwalia on 
space-time symmetries and their importance for hadron dynamics
helped clarifying various aspects of the RS classification scheme
and are kindly appreciated.
 
Work supported by CONACyT Mexico.

\def\Nucl{Nucl.\ }
\def\Phys{Phys.\ }
\def\Rev{Rev.\ }
\def\Lett{Lett.\ }
\def\PL{\Phys\Lett}
\def\PLB{\Phys\Lett B}
\def\NP{\Nucl\Phys}
\def\NPA{\Nucl\Phys A}
\def\NPB{\Nucl\Phys B}
\def\NPBS{\Nucl\Phys (Proc.\ Suppl.\ )B}
\def\PR{\Phys\Rev}
\def\PRL{\Phys\Rev\Lett}
\def\PRC{\Phys\Rev C}
\def\PRD{\Phys\Rev D}
\def\RMP{\Rev  Mod.\ \Phys}
\def\ZP{Z.\ \Phys}
\def\ZPA{Z.\ \Phys A}
\def\ZPC{Z.\ \Phys C}
\def\AOP{Ann.\ \Phys}
\def\PRep{\Phys Rep.\ }
\def\ANP{Adv.\ in \Nucl\Phys Vol.\ }
\def\PTP{Prog.\ Theor.\ \Phys}
\def\PTPS{Prog.\ Theor.\ \Phys Suppl.\ }
\def\PL{\Phys \Lett}
\def\JPF{J.\ Physique}
\def\FBSS{Few--Body Systems, Suppl.\ }
\def\IJMP{Int.\ J.\ Mod.\ \Phys A}
\def\NuCi{Nuovo Cimento~}

\end{document}